\documentclass[aps,prl,twocolumn,superscriptaddress]{revtex4}

\usepackage{CJK}

\usepackage{graphicx}

\usepackage{dcolumn}

\usepackage{bm}
\usepackage[normalem]{ulem}
\usepackage{color}

\usepackage{epstopdf}

\newcommand{\bx}{\mathbf{x}}

\newcommand{\bv}{\mathbf{v}}

\newcommand{\br}{\mathbf{r}}

\newcommand{\bff}{\mathbf{f}}
\newcommand{\bu}{\mathbf{u}}

\newcommand{\bw}{\mathbf{w}}

\newcommand{\bs}{\mathbf{s}}

\newcommand{\bof}{\mathbf{f}}

\newcommand{\sep}{ \ \ \ , \ \ \ }

\newcommand{\beq}{\begin{equation}}
\newcommand{\eeq}{\end{equation}}
\newcommand{\beqn}{\begin{eqnarray}}
\newcommand{\eeqn}{\end{eqnarray}}
\newcommand{\pp}{\partial}
\newcommand{\dd}{{\rm d}}
\newcommand{\ee}{{\rm e}}

\newcommand{\fig}{Fig.\ }

\newcommand{\cO}{{\cal O}}

\newcommand{\la}{\langle}

\newcommand{\ra}{\rangle}

\newcommand{\vnab}{{\bf \nabla}}

\begin{document}

\begin{CJK*}{UTF8}{gbsn}

\title{
Moving, reproducing, and dying beyond Flatland: Malthusian flocks in dimensions $d>2$}
\author{Leiming Chen (陈雷鸣)}
\email{leiming@cumt.edu.cn}
\affiliation{School of Physical science and Technology, China University of Mining and Technology, Xuzhou Jiangsu, 221116, P. R. China}
\author{Chiu Fan Lee}
\email{c.lee@imperial.ac.uk}
\affiliation{Department of Bioengineering, Imperial College London, South Kensington Campus, London SW7 2AZ, U.K.}
\author{John Toner}
\email{jjt@uoregon.edu}
\affiliation{Department of Physics and Institute for Fundamental Science, University of Oregon, Eugene, OR $97403$}

\begin{abstract}
We show that ``Malthusian flocks" -- i.e., coherently moving collections of self-propelled entities (such as living
creatures) which are being ``born"  and ``dying" during their motion -- belong to a new universality class in spatial
dimensions $d>2$. We calculate the universal exponents and scaling laws of this new universality class to
$O(\epsilon)$ in an $\epsilon=4-d$ expansion, and find these are different from the ``canonical" exponents
previously conjectured to hold for ``immortal" flocks (i.e., those without birth and death) and shown to hold for
incompressible flocks  in $d>2$. Our expansion should be quite accurate in $d=3$, allowing precise quantitative comparisons between our theory, simulations, and experiments.
\end{abstract}
\maketitle
\end{CJK*}

Two of the most important phenomena that distinguish biological from equilibrium systems are: spontaneous motion, and reproduction (along with its inevitable companion, death). The effects of motion
\cite{Active1,Active2,Active3,Active4},  and of birth and death \cite{tissue} separately  have been intensely studied in the field of
``Active matter". Much less is known about the interplay between the two when both are present.

The
presence of collective motion with a non-zero average velocity (``flocking") leads
to
a number of extremely unusual collective behaviors. Among these is
long-ranged orientational order in spatial
dimension $d=2$ \cite{Vicsek,TT1,Chate1,Chate2}, and the breakdown of linearized hydrodynamics that occurs in
many of its ordered phases \cite{TT1,TT3,birdrev}.

 The aforementioned properties
only occur for a particular symmetry of the  system and the state it is in; i.e., what ``phase" it is in.
 The phase that exhibits those properties is the ``polar ordered fluid" phase, which we will hereafter
refer to as a ``flock". This is a phase of active (i.e., self-propelled) particles in which the {\it only} order is the alignment of the particles' directions of motion.

 Most of the past work \cite{Vicsek,TT1,Chate1,Chate2,TT3,birdrev} on
 flocks has focused on systems with number conservation,
 which we will hereafter refer to as ``immortal flocks";  that is, they have ignored birth and death. For such systems, the local number density $\rho$ of ``flockers" (i.e., self-propelled particles) is a hydrodynamic variable. This considerably complicates the hydrodynamic theory; in particular, it leads
to six additional relevant non-linearities in the equations of motion (EOM) \cite{tonerPRE2012}, rendering the problem
intractable.

One system about which more can be said is incompressible flocks \cite{chen_njp18,chen_nc_2016}, i.e.,
 flocks in which the density is fixed, either by an infinitely stiff
equation of state, or by long-ranged forces. For these systems, it is possible to obtain exact exponents for all
spatial dimensions; as for
compressible
immortal flocks, these prove to be anomalous
(i.e., the breakdown of linearized hydrodynamics) for spatial dimensions $d$ in the range $2\le d\le4$.
Specifically, there are three universal exponents characterizing the hydrodynamic behavior of these systems.
One is  the ``dynamical exponent" $z$, which gives the scaling of hydrodynamic time scales
$t$ with length scales $L_\perp$ perpendicular to the mean direction of flock motion
(i.e., the direction of the average velocity $\left<\bv\right>$); that is, $t(L_\perp)\propto L_\perp^z$. Likewise, the growth of
length scales $L_\parallel$ {\it along} the direction of flock motion with
$L_\perp$ is characterized by an ``anisotropy exponent" $\zeta$ defined via  $L_\parallel(L_\perp)\propto L_\perp^\zeta$. Finally,  fluctuations $\bu_\perp$ of the local velocity perpendicular to its mean direction define a ``roughness exponent" $\chi$ via   $\bu_\perp\propto L_\perp^\chi$. For incompressible flocks without momentum conservation, as is appropriate for motion over a frictional substrate which acts as a momentum sink, these exponents are given by
\beq
z={2(d+1)\over5} \,\,, \,\,\, \zeta={d+1\over5}\,\,\,,\,\, \chi={3-2d\over5} \,\,,
\label{canon}
\eeq
for spatial dimensions satisfying $2 < d\le4$ \cite{chen_njp18},

The exponents (\ref{canon}) were originally asserted \cite{TT1,TT3} to hold for compressible
 immortal flocks, but this was later shown to be incorrect \cite{tonerPRE2012}, due to the presence of  the aforementioned six density non-linearities.

In this paper, we will  study the interplay of motion with birth and death by considering so-called  ``Malthusian flocks"
\cite{toner_prl12}; that is,
 flocks in which  {\it flocker number} is {\it not} conserved. Nor is momentum, due to the presence of a frictional substrate. Such systems are realizable in experiments on, e.g., growing bacteria colonies and cell tissues, and ``treadmilling''
molecular motor propelled biological macromolecules in
a variety of intracellular structures, including the cytoskleton,
and mitotic spindles, in which molecules are
being created and destroyed as they move on a frictional substrate.

In addition to describing biological and other active systems, our model for Malthusian flocks  may also be viewed as  a generic non-equilibrium $d$-dimensional $d$-component spin model  in which the spin vector space $\bs(\br)$ and the coordinate
space $\br$ are treated on an equal footing, and couplings between the two are allowed. In particular, terms like $\bs \cdot \nabla \bs$ and $(\nabla \cdot \bs)\bs$,
will be present in the EOM that describes such a generic non-equilibrium system. As a result, the fluctuations in the system can propagate spatially in a spin-direction-dependent manner, but the spins themselves are not moving. Therefore, there are no density fluctuations and the only hydrodynamic variable is the spin field,
the EOM for which is exactly the same as the one we derive here for a Malthusian flock, with spin playing the role of the velocity field.

 For Malthusian flocks, exact exponents can be obtained in $d=2$ \cite{toner_prl12}, and they again take on the  ``canonical" values
  \beq
z={6\over5} \,\,, \,\,\, \zeta={3\over5}\,\,\,,\,\, \chi=-{1\over5} \,\,.
\label{canond=2}
\eeq

 Overall, the theoretical situation is still quite unsatisfactory: we only have the scaling laws for
 flocks if they either are incompressible (which requires either infinitely strong, or infinitely ranged, interactions), or in $d=2$.
And in the cases in which we {\it do} know the exponents, their values are either
the canonical ones (\ref{canon}) \cite{toner_prl12,chen_njp18}, or those from the (1+1)-dimensional KPZ model \cite{chen_nc_2016}.

It would clearly be desirable to find the scaling laws and exponents of some  compressible three dimensional
 flocks, and to see if, as for incompressible
 flocks,
they are also given by the canonical values (\ref{canon}).

In this paper, we do so for Malthusian flocks in $d>2$. Specifically, we study these systems in an $\epsilon=4-d$ expansion. We find that they belong to a new universality class which does {\it not} have the canonical exponents (\ref{canon}). Instead, we find, to leading order in $\epsilon$,
\beqn
&&z = 2 -  \frac{6 \epsilon}{11} +\cO(\epsilon^2) \,,\label{eps1}\\
&&\zeta  = 1 -  \frac{3 \epsilon}{11} +\cO(\epsilon^2)   \,,\label{eps2}\\
&&\chi = -1 + \frac{6 \epsilon}{11} +\cO(\epsilon^2)\, ,
\label{eps3}
\eeqn
which the interested reader can easily check do {\it not} agree with the ``canonical" values (\ref{canon}) near $d=4$ (i.e., for small $\epsilon$).

Recent numerical work \cite{ginpreprint} on the far more difficult problem of
compressible immortal flocks has found that the canonical exponents
 (\ref{canon}) do not apply for that problem either. This is very consistent with  our results here for the  three
 dimensional Malthusian flocks problem, although obviously, the precise values of the exponents will be different in
 compressible immortal flocks. Indeed, they certainly are in $d=2$, where
the canonical results (\ref{canon}) {\it do} apply for Malthusian flocks, but do not, according to the simulations of \cite{ginpreprint}  for number conserving flocks.

 We have also estimated the exponents in $d=3$ by applying the one-loop (i.e., lowest order in perturbation theory) perturbative renormalization group recursion relations in arbitrary spatial dimensions. This approach, although strictly speaking an uncontrolled approximation,
not only recovers the exact linear order in the $\epsilon$-expansion results (\ref{eps1}-\ref{eps3}), but it also recovers the exact results (\ref{canond=2}) in $d=2$. Thus, while uncontrolled, this approach should provide a very effective interpolation formula for $d$ between $2$ and $4$,
that should be quite accurate (indeed, probably more accurate than the $\epsilon$ expansion) in $d=3$.

\begin{figure}
	\begin{center}
		\includegraphics[scale=.6]{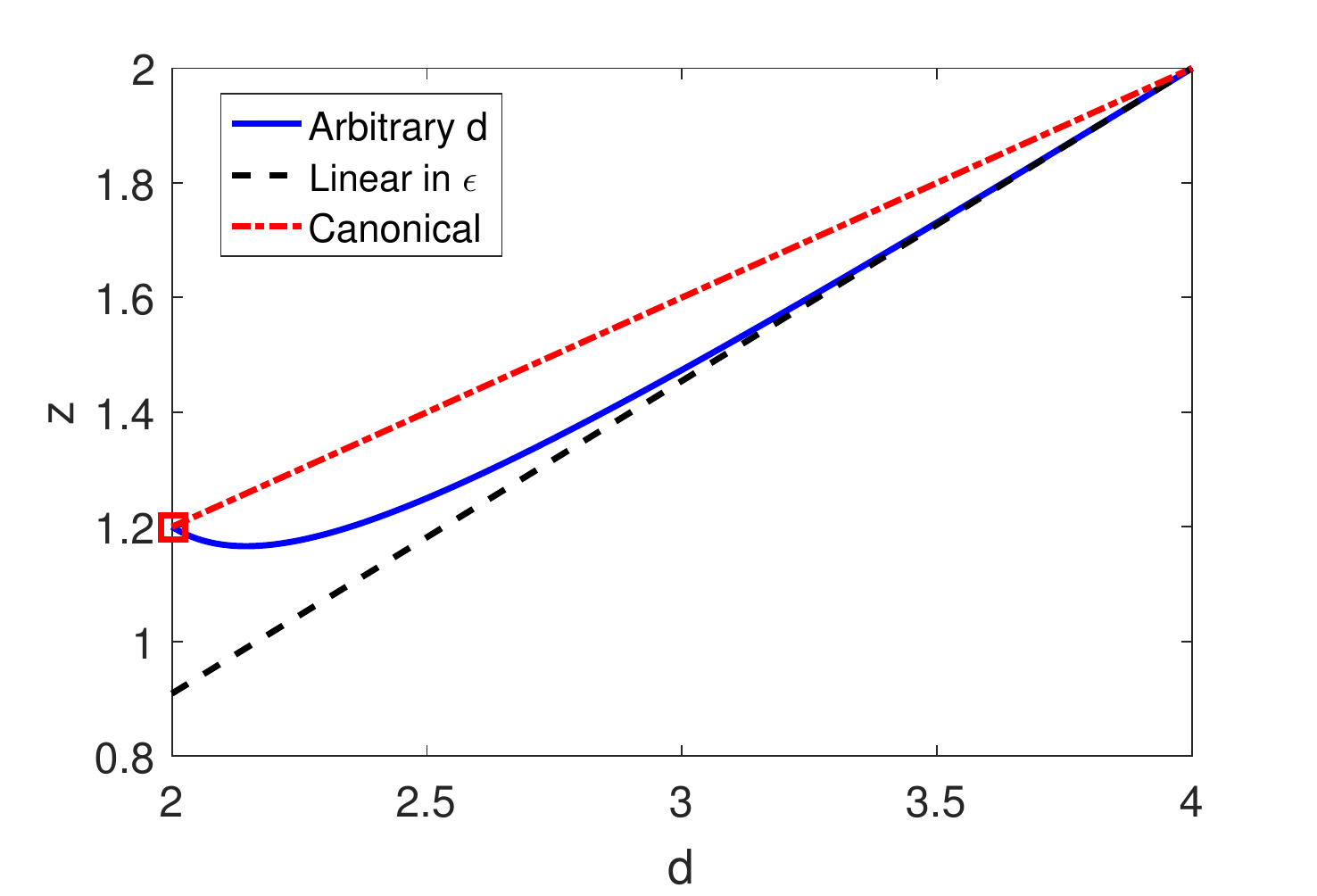}
	\end{center}
	\caption{The dynamic exponent $z$ as a function of the spatial dimension $d$. The result (\ref{eps1}) based on the $\epsilon$-expansion method to $\cO(\epsilon)$ is shown by the dashed black line, while the extrapolation (\ref{unczeta}) to arbitrary $d$ based on our one-loop result is shown in the blue curve, which converges to the known exact value (red square) in $d=2$. The canonical dynamic exponent (\ref{canon}) is shown by the dash-dotted line in red.}
	\label{fig:exponent}
\end{figure}

Using this approach, we find
\beqn
\label{uncz}
&&z = 2 -  \frac{2(4-d)(4d-7)}{14d-23}=\frac{28}{19}\ , \ \ (d=3),\\
\label{unczeta}
&&\zeta  = 1 -  \frac{(4-d)(4d-7)}{14d-23}=\frac{14}{19} \ , \ \  (d=3),  \\
\label{uncchi}
&&\chi = -1 + \frac{2(4-d)(4d-7)}{14d-23}=-\frac{9}{19}\ , \ (d=3),
\eeqn
which indeed recover our $\epsilon$-expansion results near $d=4$, and the exact results (\ref{canond=2}) in $d=2$, as the readers can verify for themselves. The result (\ref{uncz}) for $z(d)$ is graphically compared with the ``canonical" and $\epsilon$ expansion results in Fig.~\ref{fig:exponent}.

These exponents govern the scaling behavior of the experimentally measurable velocity correlation function:
 \begin{eqnarray}
&&\langle\bu_{\perp}(\br,t)\cdot\bu_{\perp}({\mathbf{0},0})\rangle\nonumber\\
&\sim&\left\{
\begin{array}{ll}
r_{_\perp}^{2\chi},&|x-\gamma t|\ll r_{_\perp}^{\zeta}, |t|\ll r_{_\perp}^z\\
|x-\gamma t|^{2\chi\over\zeta},&|x-\gamma t|\gg r_{_\perp}^{\zeta}, |x-\gamma t|\gg |t|^{\zeta\over z}\\
|t|^{2\chi\over z},&|t|\gg r_{_\perp}^z, |t|\gg |x-\gamma t|^{z\over\zeta} \,,
\end{array}
\right.
\end{eqnarray}
where $\gamma$ is a system dependent speed.

{\it Equation of motion.}
The EOM for a Malthusian flock was derived in \cite{toner_prl12}. We review this derivation in detail in the associated long paper (ALP) \cite{ALP}; here, we will only briefly outline the salient points.

Our starting EOM for the velocity is exactly that of an immortal flock \cite{TT1, TT3}:
\begin{eqnarray}
&&\partial_{t}
\bv+\lambda_1 (\bv\cdot\vnab)\bv+
\lambda_2 (\vnab\cdot\bv)\bv
+\lambda_3\vnab(|\bv|^2)
 =\nonumber \\&&
U(\rho, |\bv|)\bv -\vnab P_1 -\bv
\left( \bv \cdot \vnab  P_2 (\rho,|\bv|) \right)\nonumber \\&&+   \mu_{B} \vnab
(\vnab
\cdot \bv) + \mu_{T}\nabla^{2}\bv +
\mu_{A}(\bv\cdot\vnab)^{2}\bv+\bof
\label{vEOM}
\end{eqnarray}
where all of the parameters $\lambda_i (i = 1 \to 3)$,
$U$,
$\mu_{B,T,A}$ (``A" for ``anisotropic") and the  ``pressures'' $P_{1,2}(\rho,
|\bv|)$ are, in general, functions of the number density $\rho \equiv\rho_0+\delta\rho(\br,t)$, where $\rho_0$ is the mean density,
and the magnitude $|\bv|$ of the local velocity.
We will expand  $P_{1,2}(\rho,
|\bv|)$ about $\rho_0$.
We also find that the $\rho$ and $|\bv|$ dependence of all of the other terms does not change the long-distance scaling behavior of these systems, and therefore drop it.

In (\ref{vEOM}),
$\mu_{T,A}$ must both be positive for stability, while
$U(|\bv|<v_0)>0$, and  $U(|\bv|>v_0)< 0$ in the ordered phase. This last condition insures that in the absence of fluctuations, the flock will move at a speed $v_0$.

The $\bof$ term in (\ref{vEOM}) is a random Gaussian white noise, reflecting  errors made by the flockers, with correlations:
\begin{eqnarray}
\la f_{i}(\br,t)f_{j}(\br',t')\ra
=2D\delta_{ij}\delta^{d}(\br-\br')\delta(t-t')
\label{white noise}
\end{eqnarray}
where the noise strength $D$ is a constant hydrodynamic parameter
and $i , j$ label vector components.

We now need an EOM for $\rho$. In immortal flocks, this is just  the usual continuity equation of compressible fluid dynamics. For Malthusian flocks, it must also include the effects of birth and
death. As first noted by Malthus \cite{Malthus_1789}, {\it any}  collection of entities that is reproducing and dying can only reach a non-zero steady state population density $\rho_0$ if the difference $\kappa(\rho)$ between the birth rate
and the death rate
 vanishes at some fixed point density $\rho_0$, with larger densities decreasing (i.e., $\kappa(\rho > \rho_0) < 0)$, and smaller densities increasing (i.e.,
$\kappa(\rho < \rho_0) > 0)$.

The density EOM is therefore simply
\begin{eqnarray}
\partial_t\rho +\vnab\cdot(\bv\rho)=\kappa(\rho)~~.
\label{conservation}
\end{eqnarray}

Since birth and death quickly restore the
fixed point density $\rho_0$, we will
write $\rho(\br, t)
= \rho_0 + \delta\rho(\br, t)$ and expand both sides of equation
(\ref{conservation}) to leading  order in $\delta\rho$.    This gives
\beq
\rho_0\vnab \cdot \bv \cong \kappa' (\rho_0)\delta\rho ,
\eeq
where we've dropped the $\partial_t\rho$ and $\bv \cdot \nabla \delta \rho$ terms relative to the
$ \kappa' (\rho_0)\delta\rho$ term since we're interested in the hydrodynamic limits, in which the fields
evolve extremely slowly in both space and time.
We can use this expression to eliminate $\rho$ from the EOM (\ref{vEOM}) for $\bv$.

In the ordered state (i.e.,  in which $\left<\bv (\br, t) \right>= v_0 {\bf \hat x}$, where we've chosen the spontaneously picked direction of mean flock motion as our $x$-axis),
we can
expand the velocity EOM for small departures $ \bu (\br,t) \equiv u_x(\br,t)   {\bf \hat x} + \bu_{\perp}(\br,t) $ of $\bv(\br,t) $ from uniform motion  with  velocity $v_0 {\bf \hat x}$:
\begin{eqnarray}
\bv (\br, t) = (v_0+u_x(\br,t))  {\bf \hat x} + \bu_{\perp}(\br,t) ~~,
\label{6}
\end{eqnarray}
where henceforth  $\perp$ denotes components perpendicular to the mean velocity (or, equivalently, ${\bf \hat x}$).

The $U$ term in this EOM causes the component $u_x$ of $\bu$ to quickly relax back
to a value determined by the local configuration of $\bu_\perp$. Hence, we can eliminate $u_x$ in much the same way as we just eliminated the density $\rho$. In the ALP \cite{ALP}, we show that
doing so, and
changing co-ordinates to a new Galilean frame $\br'$ moving with respect to our original frame
in the direction ${\bf \hat{x}}$ of mean flock motion at a suitably chosen speed $\gamma$ -- i.e., $\br'\equiv\br-\gamma t  {\bf \hat x}$ -- gives
\beqn
\nonumber
\pp_t \bu_\perp +\lambda (\bu_\perp \cdot \nabla_\perp)\bu_\perp &=& \mu_1 \nabla^2_\perp \bu_\perp + \mu_2 \nabla_\perp (\nabla_\perp \cdot \bu_\perp)
\\
&&+\mu_x \pp_x^2 \bu_\perp +\bff_\perp
\ ,
\label{eq:main}
\eeqn
where we have dropped the primes. Detailed expressions for the
``suitable" speed $\gamma$ and the diffusion constants $\mu_{1,2,x}$ in terms of the parameters of equation (\ref{vEOM}) are given in the ALP \cite{ALP}.

Stability of the homogeneous ordered state requires that $\mu_1$ and $\mu_x$ are positive, and $\mu_2 /\mu_1 >-1$ \cite{ALP}.

{\it Dynamic renormalization group (DRG) analysis.}
The only nonlinear term in the EOM, $\lambda (\bu_\perp \cdot \nabla_\perp)\bu_\perp$,  does not get renormalized because of the inherent pseudo-Galilean symmetry, i.e., the invariance of the EOM under the simultaneous
replacements: $\br_\perp \mapsto \br_\perp +t\lambda \bw$ and $\bu_\perp \mapsto \bu_\perp+\bw$ for any arbitrary constant vector $\bw$ parallel to the $\perp$ direction.
In $d=2$, the noise strength $D$ is unrenormalized as well, because in $d=2$, $\bu_\perp$ has
only one component (call it $y$), and, as a result,  the nonlinear term can be written as a total
derivative: $\lambda (\bu_\perp \cdot \nabla_\perp)\bu_\perp\mapsto \lambda u_y\pp_y u_y=\lambda
\pp_y u_y^2/2$. Hence, in $d=2$, this term can only generate terms that have at least one $y$
derivative. Since the noise strength has no such derivatives, it cannot, in $d=2$, be renormalized.

This argument does not work for $d>2$, where $\bu_{\perp}$ has more than one component, which makes it impossible to write $\lambda (\bu_\perp \cdot \nabla_\perp)\bu_\perp$ as a total derivative. As a result, the noise strength $D$ {\it does} get renormalized for $d>2$.

To probe what happens for $d>2$, we perform a DRG analysis \cite{FNS} on the EOM (\ref{eq:main}). Specifically, we  first average  over short wavelength degrees of freedom, and then perform the following rescaling:
\beq
\bx_\perp \mapsto  \ee^{\ell} \bx_\perp
\ , \
x \mapsto  \ee^{\zeta\ell} x
\ , \
t \mapsto  \ee^{z \ell} t
\ , \
\bu_\perp \mapsto  \ee^{\chi \ell} \bu_\perp
\ .
\eeq
Details of this calculation are given in the ALP \cite{ALP}. The resulting DRG flow equations of the coefficients to one-loop order
are
\beqn
\label{eq:D}
\frac{1}{D}\frac{\dd D}{\dd \ell}&=&
z-2\chi-d+1-\zeta +g_1G_D(g_2)
\\
\label{eq:l}
\frac{1}{\lambda}\frac{\dd \lambda}{\dd \ell}&=&
z+\chi-1
\\
\label{eq:mux}
\frac{1}{\mu_x}\frac{\dd \mu_x}{\dd \ell}&=&
z-2\zeta
\\
\label{eq:mu1}
\frac{1}{\mu_1}\frac{\dd \mu_1}{\dd \ell}&=&
z-2+ g_1G_{\mu_1}(g_2)
\\
\label{eq:mu2}
\frac{1}{\mu_2}\frac{\dd \mu_2}{\dd \ell}&=&
z-2+g_1G_{\mu_2}(g_2)
\ ,
\eeqn
where
we've defined the dimensionless couplings
\beq
g_1 \equiv \frac{D\lambda^2}{\sqrt{\mu_x \mu_1^5}} \frac{S_{d-1}}{(2\pi)^{d-1}} \Lambda^{4-d}
\sep
g_2 \equiv \frac{\mu_2 }{\mu_1}
\ ,
\label{gdef}
\eeq
where $S_{d-1}$ is the surface area of a ($d-1$)-dimensional unit sphere, and $\Lambda$ is the
ultraviolet cutoff. While we have found that the graphical correction to $\mu_x$ is zero up to one-loop order, we strongly suspect that it becomes non-zero at higher order. The quantities  $G_{D,\mu_1,\mu_2}(g_2)$ are hideous functions of the dimensionless coupling $g_2$,
we have exiled  their exact expressions  to the  ALP \cite{ALP}. All we need to know about these functions  is that $G_{\mu_1}>G_{\mu_2}$ for all $g_2$ in the  allowed range $g_2>-1$
(which is the range required for stability),  and that
\beq
\label{G's(g2=0)}
G_{D}(g_2=0)={3(d-2)\over32(d-1)}
\sep
G_{\mu_1}(g_2=0)={4d-7\over32(d-1)} \,.\label{G_Dmu}
\eeq

 Using their definitions (\ref{gdef}), we can easily obtain from the recursion relations (\ref{eq:D}-\ref{eq:mu2})
  a closed set of recursion relations for the dimensionless couplings $g_{1,2}$:
\beqn
\label{eq:G1}
\frac{1}{g_1} \frac{\dd g_1}{\dd \ell} &=&
\epsilon +g_1\left[G_D(g_2) -\frac{5}{2} G_{\mu_1}(g_2) \right]
\\
\label{eq:G2}
\frac{1}{g_2} \frac{\dd g_2}{\dd \ell} &=&
g_1\left[G_{\mu_2}(g_2) -G_{\mu_1}(g_2) \right]
\ .
\eeqn

\begin{figure}
	\begin{center}
		\includegraphics[scale=.48]{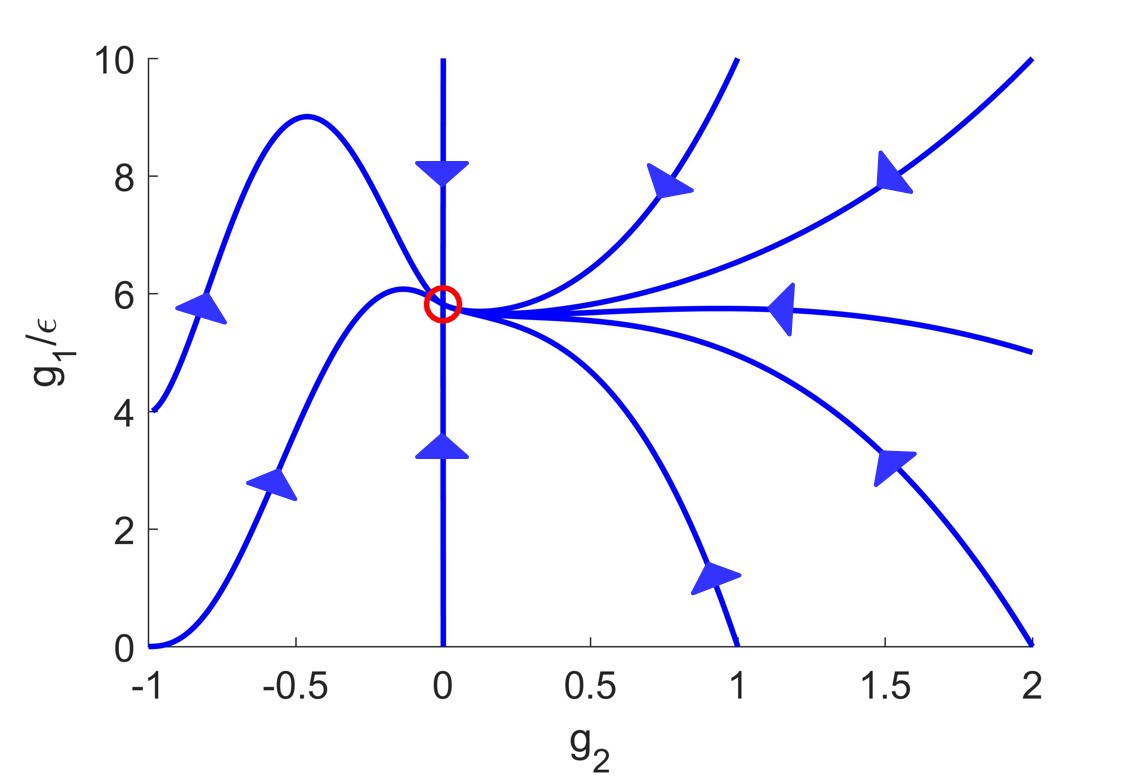}
	\end{center}
	\caption{RG flow of the coefficients $g$'s at $\epsilon = 0.1$. The stable fixed point (red circle) is at $g_1^* = 64 \epsilon/11$ and $g_2^* = 0$.
	Stability requires $g_1>0$ and $g_2>-1$.
	}
	\label{fig:RGflow}
\end{figure}

In Fig.~\ref{fig:RGflow}, we plot the RG flows of $g_{1,2}$ implied by these recursion relations.

From (\ref{eq:G2}), we see that our earlier statement that  $G_{\mu_1}>G_{\mu_2}$ for all $g_2$ implies that the only stable fixed point for $g_2$ lies at $g_2=0$, at least to the (one-loop) order to which we have worked. Setting $g_2=0$  and using
(\ref{G_Dmu}) reduces  the recursion relation (\ref{eq:G1}) for $g_1$ to
\beqn
\label{eq:G1.2}
\frac{1}{g_1} \frac{\dd g_1}{\dd \ell} &=&
\epsilon +{23-14d\over64(d-1)}g_1
\ ,
\eeqn
from which we can easily find the fixed point value $g_1^*$ of $g_1$:
\beq
\label{eq:fpgend}
g_1^* = \frac{64(d-1) }{14d-23}\epsilon +\cO(\epsilon^2)= \frac{64 }{11}\epsilon +\cO(\epsilon^2)
\ ,
\eeq
where the $\cO(\epsilon^2)$ correction in the first equality comes from higher order terms in perturbation theory that we have neglected in our one-loop approximation, while in the second equality it also incorporates corrections  from replacing the explicit $d$'s in the first equality by $4-\epsilon$.

We can now obtain  the $\epsilon$-expansion values of the scaling exponents
by inserting these fixed point values into the recursion relation (\ref{eq:mu1}) for $\mu_1$, using equation (\ref{G's(g2=0)}) with $d=4$ to evaluate $G_{\mu_1}$ at the fixed point, and choosing $z$  to keep $\mu_1$ fixed. This gives (\ref{eps1}).
Requiring that $\mu_x$ and $\lambda$ remain fixed leads to the conditions
\beq
z+\chi=1
\sep
z-2\zeta=0
\,.
\label{muxlambfix}
\eeq
Using the value of $z$ we just found in these two equations gives the $\epsilon$-expansion values (\ref{eps2},\ref{eps3}) for $\zeta$ and $\chi$.

{\it Beyond linear order in $\epsilon$.} Our results so far are based on a one-loop calculation, which accurately captures the universal behavior of the system to linear order in $\epsilon$. However, since all of our expressions for  $G_{D,\mu_1,\mu_2}$'s are for general $d$, we can extrapolate our result to arbitrary $d$  keeping only our one-loop  expressions.
Clearly, this is an uncontrolled approximation, since it ignores higher loop graphs -- that is, terms higher order than linear in $g_1$ in perturbation theory -- which will not be small if $\epsilon$ is not small, since then the fixed point value $g_1^*$ of $g_1$ is not small either. Nonetheless, as noted earlier,
this uncontrolled
approximation not only reproduces the $\epsilon$-expansion results (\ref{eps1},\ref{eps2},\ref{eps3}) near $d=4$, but also the exact results (\ref{canond=2}) in $d=2$.
Thus, this truncation, while uncontrolled, probably gives an extremely good
 interpolation
formula between the leading order $\epsilon$-expansion results near $d=4$,  and the exact results in $d=2$.

Making this truncation, one can immediately see that the argument that the only stable fixed point is at $g_2=0$ stable still applies, as does  (\ref{eq:G1.2}),
 which leads to the fixed point value of $g_1$ in terms of general $d$:
 \beq
 g_1^* = \frac{64(d-1)(4-d) }{(14d-23)}\,.
 \eeq
Using the above expression for $g_1$ and (\ref{G's(g2=0)}) for $G_{\mu_1}(g_2=0)$
in  (\ref{eq:mu1}),  and choosing $z$ to keep $\mu_1$ fixed,  gives the expression (\ref{uncz}) for the dynamical exponent $z$  for arbitrary $d$. Then using (\ref{muxlambfix}), we obtain the expressions (\ref{unczeta}) and (\ref{uncchi}) for the anisotropy exponent $\zeta$ and the roughness exponent $\chi$.
 As illustrated in \fig \ref{fig:exponent}, the numerical values obtained using this method are very similar to those from the leading order $\epsilon$-expansion results.

{\it Summary.}
Focusing on the ordered phase of a generic Malthusian flock in dimensions $d>2$, we have used a dynamic renormalization group analysis to reveal a novel universality class that describes the system's hydrodynamic properties. In particular, we estimated the scaling exponents using the conventional one-loop $\epsilon$ expansion and  an uncontrolled
one-loop approach. The latter approach recovers
the known exact result in $d=2$. In $d=3$ the
estimated values obtained by the two approaches
nearly equal, which implies our predictions are quantitatively accurate.
Our work is the first determination of the
scaling  of fluctuations away from a critical point in an active system to
require the full apparatus of the dynamical renormalization group;
in particular, the evaluation of Feynmann graphs.

\begin{acknowledgments}
LC acknowledges support by the National Science Foundation of China (under Grant No. 11874420). JT thanks  The Higgs Centre for Theoretical Physics at the University of Edinburgh for their hospitality and support while this work was in progress.
\end{acknowledgments}


\begin{thebibliography}{7}
	\expandafter\ifx\csname natexlab\endcsname\relax\def\natexlab#1{#1}\fi
	\expandafter\ifx\csname bibnamefont\endcsname\relax
	\def\bibnamefont#1{#1}\fi
	\expandafter\ifx\csname bibfnamefont\endcsname\relax
	\def\bibfnamefont#1{#1}\fi
	\expandafter\ifx\csname citenamefont\endcsname\relax
	\def\citenamefont#1{#1}\fi
	\expandafter\ifx\csname url\endcsname\relax
	\def\url#1{\texttt{#1}}\fi
	\expandafter\ifx\csname urlprefix\endcsname\relax\def\urlprefix{URl }\fi
	\providecommand{\bibinfo}[2]{#2}
	\providecommand{\eprint}[2][]{\url{#2}}
	

		\bibitem{Active1}
	S.~Ramaswamy, The mechanics and statics of active matter. Ann. Rev. Condens. Matt. Phys. {\bf 1}, 323-345 (2010).
	
			\bibitem{Active2}
 M.C.~Marchetti, J.F.~Joanny, S.~Ramaswamy, T.B.~Liverpool, J.~Prost, M.~Rao,  and R.A.~Simha, Hydrodynamics of soft active matter, {\it Rev. Mod. Phys.} {\bf 85}, 1143-1188 (2013).
 	
		\bibitem{Active3}
 C.~Bechinger, R.~Di Leonardo, H.~L\"{o}wen, C.~Reichhardt, G.~Volpe, and G.~Volpe, Active particles in complex and crowded environments, Rev. Mod. Phys. {\bf 88}, 045006 (2016).
 	
		\bibitem{Active4}
	F.~Schweitzer, {\it Brownian Agents and Active Particles: Collective Dynamics in the Natural and Social Sciences}. Springer Series in Synergetics (Springer, New York, 2003).
	
	\bibitem{tissue} J.~Ranft, M.~Basan, J.~Elgeti, J.-F.~Joanny, J.~Prost and F.~J\"{u}licher,  Fluidization of Tissues by Cell Division and Apoptosis, Proc. Natl. Acad. Sci. USA {\bf 107}, 20863 (2010); J.~Ranft, J.~Prost, F.~J\"{u}licher, and  J.-F.~Joanny,  Tissue Dynamics with Permeation, Eur. Phys. J. E {\bf 35}, 46 (2012).

	
        \bibitem{Vicsek}
T. Vicsek, A. Czir\'{o}k, E. Ben-jacob, I. Cohen, and O. Shochet, Novel type of phase transition in a system of self-Driven particles. Phys.\ Rev.\ Lett. {\bf 75}, 1226 (1995).

\bibitem{TT1}
	J.~Toner, and Y.~Tu, Long-range order in a two-dimensional dynamical XY model: how birds fly together. Phys.\ Rev.\ Lett. {\bf 75}, 4326 (1995).

\bibitem{Chate1}
G. Gr\'{e}goire and H. Chat\'{e}, Onset of Collective and Cohesive Motion. Phys.\ Rev.\ Lett. {\bf 92}, 025702 (2004).

\bibitem{Chate2}
H. Chat\'{e}, F. Ginelli, G. Gr\'{e}goire, and F. Raynaud, Collective motion of self-propelled particles interacting without cohesion. Phys.\ Rev.\ E {\bf 77}, 046113 (2008).

			
	
	%
	\bibitem{TT3}
	J.~Toner, and Y.~Tu, Flocks, herds, and schools: a quantitative theory of flocking. Phys.\ Rev.\ E {\bf 58}, 4828(1998).
	
	\bibitem{birdrev} J.\ Toner,   Y. Tu, and S.
Ramaswamy, Hydrodynamics and phases of flocks.  Ann.\ Phys.\  {\bf 318}, 170 (2005).

    \bibitem{tonerPRE2012}
    J.\ Toner, Reanalysis of the hydrodynamic theory of fluid, polar-ordered flocks.
Phys.\ Rev.\ E {\bf 86}, 031918-1-031918-9 (2012).

    \bibitem{chen_njp18}
	L.~Chen, C.~F.~Lee, and J.~Toner, Incompressible polar active fluids in the moving
    phase in dimensions $d>2$. New J.~Phys. {\bf 20}, 113035 (2018).
	
    \bibitem{chen_nc_2016}	
    L.~Chen, C.~F.~Lee, and J.~Toner, Mapping two-dimensional polar active fluids to
    two-dimensional soap and one-dimensional sandblasting. Nat.~Commun. {\bf 7}, 12215 (2016).

    \bibitem[{\citenamefont{Toner}(2012)}]{toner_prl12}
	\bibinfo{author}{\bibfnamefont{J.}~\bibnamefont{Toner}},
	\bibinfo{title}{Birth, Death, and Flight: A Theory of Malthusian Flocks}.
    \bibinfo{journal}{Physical Review Letters} \textbf{\bibinfo{volume}{108}},
	\bibinfo{pages}{088102} (\bibinfo{year}{2012}),
	\urlprefix\url{http://dx.doi.org/10.1103/PhysRevLett.108.088102}.

 \bibitem{ginpreprint} B. Mahault, F. Ginelli, and H. Chate, Quantitative Assessment of the Toner and Tu Theory of Polar Flocks,  ArXiv:1908.03794.

    \bibitem{ALP} The accompanying long paper titled  ``A novel nonequilibrium state of matter: a $4-\epsilon$ expansion study of Malthusian flocks''.





	
	   \bibitem{Malthus_1789}
T.\ R.\ Malthus, An Essay on the Principle of Population, edited by J. Johnson (St. Paul's Churchyard, London, 1798).



\bibitem{FNS}
D. Forster, D. R. Nelson and M. J. Stephen, Large-distance and long-time properties of a randomly
stirreduid. Phys. Rev. A {\bf 16} 732 (1977).

%
%
%
%
%
	
	
	
	


%
%
%
%
%
%
	
\end{thebibliography}
\end{document}